\documentstyle[prb,twocolumn,aps,epsf]{revtex}
\begin{document}
\draft
\twocolumn[\hsize\textwidth\columnwidth\hsize\csname@twocolumnfalse%
\endcsname

\title{Mean Free Path and Energy Fluctuations in Quantum Chaotic Billiards}

\author{E. Louis$^1$, E. Cuevas$^{1,2}$, J. A. Verg\'es$^3$ and
M. Ortu\~no,$^2$}

\address{$^1$Departamento de F{\'\i}sica Aplicada, Universidad de Alicante,
\\ Apartado 99, E-03080 Alicante, Spain.}

\address{$^2$Departamento de F{\'\i}sica,
Universidad de Murcia, E-30071 Murcia, Spain.}

\address{$^3${I}nstituto de Ciencia de Materiales de Madrid, \\
Consejo Superior de Investigaciones Cient{\'\i}ficas,
Cantoblanco, E-28049 Madrid, Spain.}

\date{August 7, 1996}

\maketitle

\begin{abstract}
The elastic mean free path of carriers in a recently introduced model
of quantum chaotic billiards in two and three dimensions is calculated.
The model incorporates
surface roughness at a microscopic scale by randomly choosing the atomic
levels at the surface sites between $-W/2$ and $W/2$. 
Surface roughness yields a mean free path $l$ that decreases as
$L/W^2$ as $W$ increases, $L$ being the linear system size.
But this diminution ceases when the surface layer begins to decouple from
the bulk for large enough values of $W$, leaving more or less unperturbed
states on the bulk. Consequently, 
the mean free path shows a minimum of about $L/2$
for $W$ of the order of the band width.
Energy fluctuations reflect the behavior of the mean free path.
At small energy scales, strong level correlations manifest themselves by
small values of $\Sigma^2(E)$ that are close to Random Matrix Theory (RMT)
in all cases. At larger energy scales,
fluctuations are below the logarithmic behavior of RMT for
$l > L$, and above RMT value when $l < L$. 
\end{abstract}

]
\narrowtext
\section{Introduction}

Quantum systems in which the scattering of carriers by bulk impurities
or defects is negligible are traditionally considered as  ballistic.
An interesting example of these systems are the so called billiards:
a region of space in which all scattering occurs at the boundary.
The ballistic character of these systems is usually taken
for granted, even in the case that their classical analogs behave
chaotically \cite{Na94}. This is so despite of the fact that in wave
mechanics the localized character of the scattering centers does not
guarantee that their effects be also localized.
On the other hand, experimental data for the mean free path of
microcavities do always correspond to the bulk material from
which the cavity was produced. For instance, Chang {\it et al.}\cite{exp}
give a transport mean free path of the bulk material which is ten times 
larger than the cavity diameter.

In this work, we characterize quantum billiards by the standard magnitude used
in transport studies, i.e., by its mean free path.
Surprisingly, the calculated elastic mean free path $l$ is smaller than the
linear size of the system for physically relevant values of surface roughness.
Afterwards, the consequences of such small values of
$l$ are studied on measurable magnitudes like the long--range
energy correlations within the spectrum. Our calculations are done for
a model of quantum chaotic billiard recently proposed by
the authors\cite{CL96a,VC96}.
The model considers a system which is perfect all throughout the bulk,
whereas its surface shows roughness at a microscopic scale.
Bulk disorder is not considered since, as remarked above, 
experimental mean free paths of the bulk systems are usually much 
larger than cavity diameters.
An efficient implementation of this idea was developed in our previous
work \cite{CL96a}: a tight-binding Hamiltonian with a single atomic level
per lattice site in which the energies of the atomic
orbitals at the surface sites are chosen at random. 
We proved that, in the thermodynamic limit, this model shows the 
accepted signs of quantum chaotic behavior \cite{Gu90,BF81,Bo91}:
an energy spectrum characterized by a Wigner--Dyson distribution
of nearest--level spacings, and typical scars in the probability
amplitude of wavefunctions. 

Section II introduces the quantum billiard model. The elastic mean free path
is calculated in two complementary ways in Section III.
Long--range energy fluctuations, which are known to be a clear
hallmark of diffusive or ballistic behaviors \cite{AS86,AG95,CL96b},
are analyzed in Section IV.
The paper ends with a Section that collects our main results.

\section{Quantum Billiard Model}

Our quantum billiard model is defined by the following
tight--binding Hamiltonian:

\begin{equation}
\widehat H = \sum_{i \in S} \omega_i \hat c_i^{\dagger} \hat c_i
+ \sum_{\langle ij \rangle } t_{ij} \hat c^{\dagger}_i \hat c_j\;,
\label{hamiltonian}
\end{equation}

\noindent
where the operator $\hat c_i$ destroys an electron on site $i$, and
$t_{ij}$ is the hopping integral between sites $i$ and $j$ (the symbol
$\langle ij\rangle $ denotes that the sum is restricted to nearest neighbor 
sites). We take $t_{ij}=t=-1$ and consider square and cubic lattices for
2D and 3D systems, respectively.
The energy $\omega_i$ of a given atomic level
at the surface ($S$) site $i$, is randomly chosen between $-W/2$ and $W/2$, 
whereas other sites have a constant energy equal to zero.
Calculations have been carried out for squares of side up to $L = 130$
and for cubes of side up to $L=20$, where $L$ is given in units of the
lattice constant.
The Schwarz algorithm for symmetric band matrices\cite{Sc68} was used to
compute the whole spectrum.

Note that this model, in contrast with standard billiards \cite{Gu90},
has two length scales: the system size $L$ and the lattice constant $a$.
Thus, even in the macroscopic limit ($L/a \rightarrow \infty$) microscopic
roughness remains, and is felt by quantum particles, i.e., particles
characterized by a wavelength of the order of $a$. 

\section{Elastic Mean Free Path}

The ratio $l/L$, $l$ being the mean free path, is the parameter that
controls whether a system is diffusive or ballistic and so the type of
fluctuations \cite{AS86,AG95,CL96b}. The standard definition of the
elastic mean free path is \cite{Ec83,Sh95}:

\begin{equation}
l_{\bf k} \equiv v_{\bf k}\tau_{\bf k}=\frac{\hbar v_{\bf k}}
{2|{\rm Im}\Sigma({\bf k},{\bf k};\epsilon_{\bf k})|}\;,
\label{ldef}
\end{equation}

\noindent
where $\tau_{\bf k}$ is the relaxation time and $v_{\bf k}$ the
velocity of a state with momentum ${\bf k}$.
We calculate the selfenergy induced by disorder in an ordered state of
momentum ${\bf k}$ and energy $\epsilon_{\bf k}$ from Dyson equation:

\begin{equation}
\widehat{\Sigma}(\epsilon_{\bf k}) \equiv 
{\langle \widehat{G}(\epsilon_{\bf k}) \rangle}^{-1} -
\widehat{G}^{-1}_0(\epsilon_{\bf k})\;,
\label{autoenergia}
\end{equation}

\noindent
where $\widehat{\Sigma}$ is the selfenergy operator,
$\widehat{G}_0$ the unperturbed Green function, and
$\langle \widehat{G} \rangle$ the perturbed Green function
averaged over disorder realizations.
Then, in order to calculate the matrix element
$\Sigma({\bf k},{\bf k}; \epsilon_{\bf k})$, we only have to determine
the inverse of the averaged perturbed Green function in the
{\bf k}--space basis using its site representation:

\begin{equation}
\langle {\bf k} | {\langle \widehat{G}(\epsilon_{\bf k}) \rangle}^{-1} |
{\bf k} \rangle
=\sum_{i,j} \psi_{\bf k}({\bf r}_i) \psi_{\bf k}({\bf r}_j)
\langle G(i,j;\epsilon_{\bf k}) \rangle ^{-1}\;,
\end{equation}

\noindent
where $\psi_{\bf k}({\bf r}_i)$ are the wave functions of the ordered cluster:

\begin{mathletters}
\begin{equation}
\psi_{\bf k}^{2D}({\bf r}_i)=\frac {2}{L+1} \sin(k_x x_i)
\sin(k_y y_i)\;,
\end{equation}
\begin{equation}
\psi_{\bf k}^{3D}({\bf r}_i)=\frac {2\sqrt{2}}{(L+1)^{3/2}}
\sin(k_x x_i) \sin(k_y y_i) \sin(k_z z_i)\;.
\end{equation}
\end{mathletters}

It should be noted that non--diagonal elements in Eq.(4) are essential
in this case, due to the strong breaking of translational invariance
introduced by surface disorder. On the other hand, we have checked that
a calculation of the selfenergy due to bulk disorder without the
incorporation of non--diagonal elements gives very accurate
results\cite{Ec83,Sh95}.

The qualitative behavior of $l$ can be obtained by second-order perturbation
theory\cite{Ec83}. Averaged real--space matrix elements of the selfenergy
operator are approximately given by:

\begin{equation}
\Sigma(i,j;\epsilon_{\bf k}) \approx {\delta_{i j}} \langle\omega_i^2\rangle 
G_0(i,i;\epsilon_{\bf k})\;,
\end{equation}

\noindent
for those sites $i$ for which the diagonal energy $\omega_i$ fluctuates,
i.e., for the surface sites. Using the basis given by Eq.(5) to transform
to {\bf k}--space and assuming a constant density of states
($1/(4d)$, being $d$ the dimension of the physical space)
to get a rough approximation for the imaginary part
of the diagonal elements of the unperturbed Green function
(${\rm Im}[G_0(i,i;\epsilon_{\bf k})] \sim {\pi / (4d)}$),
we obtain:

\begin{equation}
{\rm Im} \Sigma({\bf k},{\bf k};\epsilon_{\bf k}) \approx
{{\pi \over 24} {W^2 \over L}} \; .
\end{equation}

\noindent
Substituting the selfenergy in Eq.(\ref{ldef}) for this approximate value
and using the 2D dispersion relation of the band
$e_{\bf k}=-2[\cos(k_x)+\cos(k_y)]$) to get an explicit equation for
$\hbar  v_{\bf k}$, we finally have:

\begin{equation}
l_{\bf k} \approx \frac{24}{\pi} \frac{L}{W^2} 
\sqrt{\sin^2(k_x)+\sin^2(k_y)} \; .
\label{l}
\end{equation}

\noindent
The final expression for $l$ only uses dimensionless magnitudes, i.e.,
disorder energies $W$ are measured in units of the hopping energy $t$,
lengths in units of the lattice constant $a$, and ${\bf k}$ in units
of $a^{-1}$.  The corresponding expression for 3D has an additional term
within the square root.

\begin{figure}
\begin{picture}(235,260) (50,-5)
\epsfbox{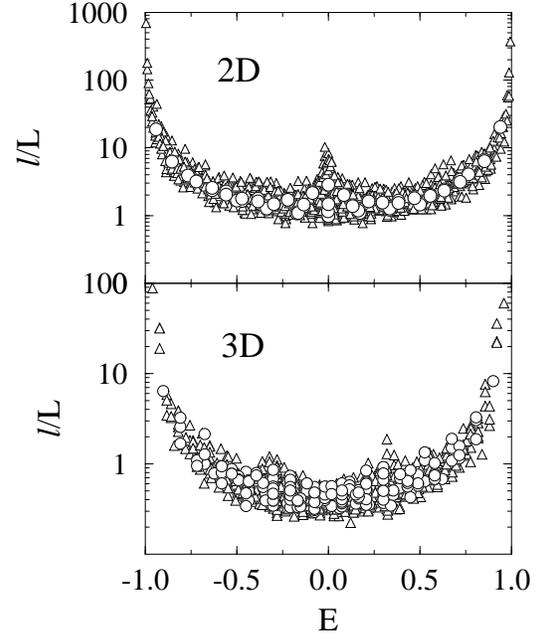}
\end{picture}
\caption{
Mean free path (divided by the system size $L$) as a function of
the energy (scaled by half the band width, 4 and 6 in 2D and 3D,
respectively) for the
chaotic billiards discussed in this work. In the 2D case the results
correspond to $L$=8 and 32 (circles and triangles) and $W=2$, and in the
3D case to $L=6$ and 10 (circles and triangles) and $W=4$. The fact that
the data for the two sizes collapse indicates that $l \propto L$ (see text).
\label{fig.mfp_2D_3D}}
\end{figure}

When second--order perturbation theory is used to get an approximate
value of the mean free path induced by bulk disorder,
a rather good approximation\cite{Ec83,Sh95} is obtained due to the
homogeneous character of both the perturbation and the real--space Green
function. Nevertheless, when the same scheme is applied to billiards,
it gives the correct $L$ and $W^{-2}$
dependences but fails to describe the variation of
$l$ within the band (see our numerical results shown in
Fig.\ref{fig.mfp_2D_3D}).

Comparison of the mean free path of a billiard (Eq.(\ref{l})) with the
result obtained for bulk disorder\cite{Ec83} reveals an essential
feature of the present model: whereas for bulk disorder $l$ is
independent of the system size $L$, in the present case it depends linearly
on $L$.
This result is a consequence of having the scattering centers located at
the surface, an essential feature of billiards. Therefore, ballistic
($l \geq L$) or even clean behavior ($l \gg L$)
could naively be expected over the whole band for any
value of $W$. Our results will show that this is not the case for
physically relevant values of $W$.

Fig.\ref{fig.mfp_2D_3D} shows typical numerical results for the mean free
path in 2D and 3D.
Note that $l$ is indeed proportional to the system size $L$ 
in agreement with the approximate result reported in Eq.(\ref{l}).
The dependence
of the mean free path on the energy reflects the nature of the eigenvalues
discussed in our previous works \cite{CL96a,VC96}.
In 2D, $l$ increases near the edges and at the center of the band, due to
the presence of quasi-ideal states in these energy regions\cite{quasi}.
Quasi-ideal states do also appear in three dimensions near
the band edges and this fact is also
reflected in the behavior of the mean free path. The energy dependence
of $l$ is significantly weaker than in the case of bulk disorder.
Actually,
$l$ is almost constant except near the band edges and the band center.

Eq.(\ref{l}) shows that the mean free path
monotonically decreases with $W$. However, we note
that the definition of the mean free path (Eq.(\ref{ldef})) is intrinsically
perturbative. As a consequence, the results for large $W$ cannot 
be trusted. In fact in the limit of large $W$ the surface decouples from 
the bulk, leaving a fully ordered cluster of linear size $L-2$.
Consequently, the mean free path {\it of the inner part}
should tend to infinity.
In order to handle this problem, we have calculated the mean
free path in the large $W$ limit by considering the perturbation
of the ordered states of the $(L-2)^d$ cluster due to the
addition of the surface layer.

Firstly,
the averaged Green function of the whole cluster of lineal size $L$
is calculated for the eigenenergies of the smaller cluster of linear
size $L-2$. Secondly, the averaged Green function of the inner cluster is
obtained using a projection operator $\widehat{P}$,
and finally, the selfenergy operator is defined by means of an equation
quite similar to Eq.(\ref{autoenergia}):

\begin{equation}
\widehat{\Sigma}(\epsilon_{\bf k}) \equiv
\left [ \widehat{P}^{\dagger} \big \langle \widehat{G}(\epsilon_{\bf k})
\big \rangle \widehat{P} \right ]^{-1} -
\widehat{G}^{-1}_0(\epsilon_{\bf k})\;.
\label{autoenergia2}
\end{equation}

Since Eq.(\ref{autoenergia2}) is an operator definition, any basis can
be used in the numerical calculation.
We use the tight--binding basis in which the
projector $\widehat{P}$ is particularly simple. Thus, the diagonal
elements of the selfenergy in the {\bf k}--space basis are given by:

\begin{equation}
\langle {\bf k} |\widehat{\Sigma}(\epsilon_{\bf k})| {\bf k} \rangle
=\sum_{i,j} \psi_{\bf k}({\bf r}_i) \psi_{\bf k}({\bf r}_j)
\Sigma(i,j;\epsilon_{\bf k})\;.
\end{equation}

\begin{figure}
\begin{picture}(235,250) (-60,-40)
\epsfbox{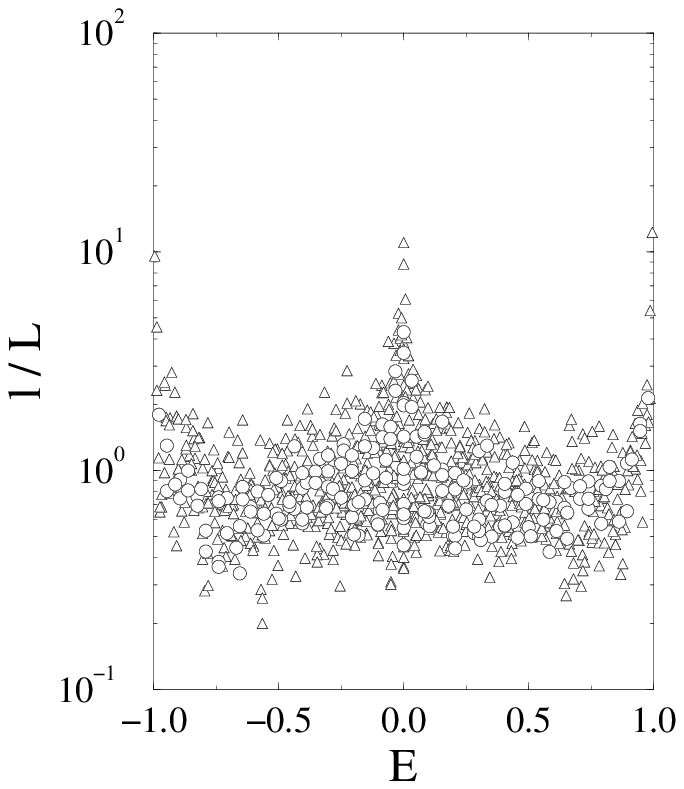}
\end{picture}
\caption{
Mean free path (divided by the system size $L$) as a function of
the energy (scaled by half the band width) for the
chaotic billiards discussed in this work.
Results correspond to $L$=16 and 32 (circles and triangles) and $W=10$.
The fact that
the data for the two sizes collapse indicates that $l \propto L$ (see text).
\label{fig.mfp_red}}
\end{figure}

This alternative perturbation theory provides us with a second value
of the mean free path. Of course, the larger value is the only one which is
physically relevant (perturbation theory logically starts from the less
perturbed system). Although we
have not found an analytic approximation for this second
value of $l$, our numerical results show that $l$ is indeed linearly
dependent on $L$ (see Fig.\ref{fig.mfp_red}). Apart from some features that
can be related to singularities of the density of states, this new value of
$l$ behaves in the same way as the standard value: it shows considerable
fluctuations around a mean value that is almost constant within the band.
Similar results have been found for three dimensional billiards.

\begin{figure}
\begin{picture}(235,250) (-60,-40)
\epsfbox{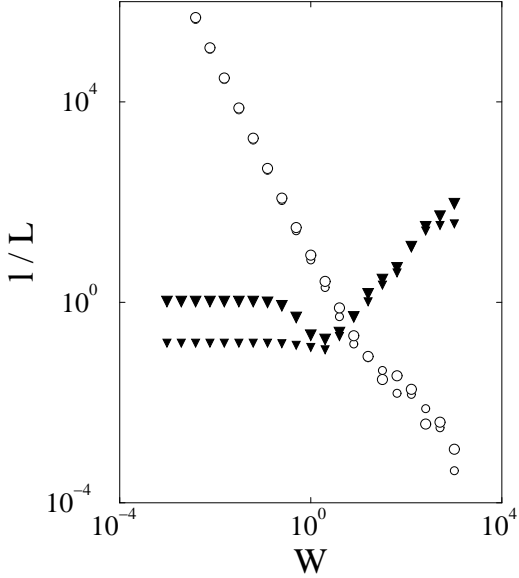}
\end{picture}
\caption{
Mean free path $l$ divided by $L$ as a function of $W$
for the quantum chaotic billiards investigated in this work.
Standard approach results are given by circles whereas our alternative
calculation is represented by down triangles.
Small symbols correspond to $L=16$ and large symbols to $L=32$.
The results correspond to an average over an energy window of 
0.2 around an energy equal to $-2$.
\label{fig.mfp_scaW}}
\end{figure}

Fig.\ref{fig.mfp_scaW} shows the scaling behavior of both values of $l$ with
disorder.
The relevant (larger) values of $l$ closely follow an $W^{-2}$ law for
small $W$ values while the scaling is close to $W$ for large $W$ values.
These scaling laws do not depend on the system size.
On the contrary, adding the external shell to the cluster is a
perturbation that decreases in importance as the cluster increases
(see the behavior of triangles in the left part of Fig.\ref{fig.mfp_scaW}).
Our results show that the ratio $l/L$ can reach values as small as 0.4
close to $W=6$. We
will see in the next section, that the value of this ratio strongly
determines the behavior of long--range correlations within the energy spectrum.
Our data show that the scaling behavior given in Fig.\ref{fig.mfp_scaW} for
two dimensional billiards holds for billiards in spaces of larger dimension.
This result
is not surprising according to well--known results for bulk disorder:
mean free path is weakly dimension dependent in those regions in which
it is almost constant\cite{Sh95}.

\begin{figure}
\begin{picture}(235,160) (-0,-15)
\epsfbox{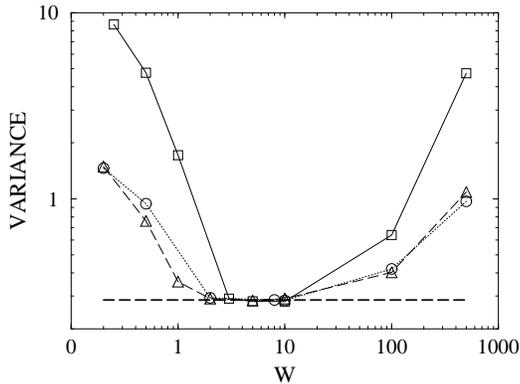}
\end{picture}
\caption{
Variance of the nearest--level spacing distributions as a function 
of the disorder parameter $W$ for the two (circles, $L$=32, triangles,
$L=90$) and three (squares, $L$=10) dimensional billiards here investigated.
The results correspond to an average over an energy window of  0.2 around
the energy $-1$. The dashed horizontal line corresponds to the GOE variance
(0.286).
\label{fig.WD}}
\end{figure}

Let us mention that although nearest-neighbors spacing statistics
has been proved to follow Wigner-Dyson statistics in our previous
work\cite{CL96a}, finite clusters only show "nice" GOE statistics for
a range of $W$ values close to the bandwidth.
Fig.\ref{fig.WD} illustrates this fact.
It shows the variance of nearest--neighbor spacing distribution
as a function of $W$ for several system sizes. As expected, chaotic
behavior is not found when disorder is either very strong or very weak.
Beyond a value of $W \approx 10$ which is of the order of the 
band width, the variance becomes larger than the value corresponding
to the Wigner--Dyson statistics (0.286), indicating that the surface
starts to decouple from the bulk. For this value of $W$, most
of the energy levels which are mainly localized on surface sites 
lie outside the band and follow Poisson statistics.
Also for small $W$ values, the value becomes larger than 0.286. This
indicates a {\it clean} behavior of our system: statistics of clean
(ordered) systems is more or less Poisson up to some energy scale\cite{casati}.
The dependence of the variance on the system size is also illustrated in
Fig.\ref{fig.WD}. When $L$ increases from
32 to 90, the lower value of $W$ below which the variance is no longer that
of GOE decreases. Further increases of $L$ would reduce to zero the widths
of the energy regions where quasi--ideal states appear, and thus, the
variance would become that of GOE through the whole band, no matter how
small $W$ would be.

\section{Energy Fluctuations}

In this section we analyze the relationship between mean free path and
measurable physical magnitudes. Although conductance was our first
candidate, finally we chose long--range energy fluctuations due to
the importance they have in analyzing quantum chaotic systems. 
Moreover, while the behavior of energy
fluctuations within the spectrum is an intrinsic property, conductance
depends on the way the system is attached to the metallic leads.

The variance of the number of levels in a given energy interval is a
well-known statistics usually referred as
$\Sigma^2(E)$ statistics\cite{BF81}.
The standard way of measuring $\Sigma^2(E)$
proceeds firstly to transform real spectra onto unfolded spectra:
each real spectrum $\{\epsilon_i\}$ is mapped onto an unfolded spectrum
$\{E_i\}$ through $E_i= {{\overline N}(\epsilon_i)}$,
where ${{\overline N}(\epsilon)}$ is the averaged number
of levels up to an energy $\epsilon$.
Secondly, the variance of the number of levels found in an interval of
fixed length $E$ is directly obtained from the unfolded spectra.
In this Section, we present results for the variance of the number of levels
obtained by a different numerical procedure. Of course, we have checked
that overall features are procedure independent, i.e., our method coincides
with the standard one when subtleties are ignored. Nevertheless, fine
details of the number variance are better given by a more direct definition.

Let us sketch the numerical procedure.
Some energy interval $[\epsilon_1,\epsilon_2]$ of the band spectrum
is chosen.  The mean number of levels within the interval is given by:

\begin{equation}
\overline N = \bigg \langle N(\epsilon_2)-N(\epsilon_1) \bigg \rangle_{W} \; ,
\end{equation}

\noindent
where $N(\epsilon)$ gives the total number of states below
energy $\epsilon$ for a generic disorder realization and averaging
over disorder configurations is done.
In the same way, the mean of the squared number of levels is given by:

\begin{equation}
\overline {N^2} = 
\bigg \langle [N(\epsilon_2)-N(\epsilon_1)]^2 \bigg \rangle_{W} \; .
\end{equation}

The variance of the number of levels contained in the energy interval
$[\epsilon_1,\epsilon_2]$ is simply:

\begin{equation}
\Sigma^2(\overline N)= \overline {N^2} - {\overline N}^2 \; .
\end{equation}

\noindent
In this way, just a value of the $\Sigma^2(E)$ function is obtained.
This sequence is repeated a large number of times for randomly selected
energy intervals until a relatively smooth value is obtained for the
variance of the number of levels {\it averaged} over energy intervals
containing the same number of levels. The last step implies averaging
the number variance over some selected region of the energy band.

Note that only fluctuations induced by disorder are taken into account
by our method. Therefore, number fluctuation is strictly zero for an
isolated spectrum. On the other hand, the standard method can be used
even in this situation. The variance is defined by:

\begin{equation}
\Sigma^2(E) = [N(E_0+E)-N(E_0)-E]^2 \; .
\end{equation}

The subtlety within this definition is the way followed for unfolding
just one spectrum.
Sometimes, the asymtotic form of $N_{\epsilon \rightarrow \infty}(\epsilon)$
(Weyl formula)\cite{Gu90}
is used even in the lower part of the spectrum and number fluctuations
are measured relative to $N_{\epsilon \rightarrow \infty}(\epsilon)$.
Most frequently, a large number of levels is used to measure the average
level spacing and then fluctuations relative to the mean value are measured
in smaller intervals. In this way, it has been proved that crystalline
spectra are uncorrelated to some energy extent,
and consequently, $\Sigma^2(E)$ statistics is Poisson\cite{casati}:

\begin{equation}
\Sigma^2_{\rm Poisson}(E)=E \; .
\label{Poisson}
\end{equation}

Let us remark that our procedure is quite appropriate for billiard
ensembles as it is our case (surface roughness is random and several
samples can be generated for a given value of surface disorder $W$).
On the other hand, it gives no fluctuations for conventional billiards (just
one spectrum).

We have checked that our numerical procedure recovers both the analytical
results for random series of energy levels (see Eq.(\ref{Poisson}))
and for the eigenvalue series of matrices belonging to the Gaussian
Orthogonal Ensemble (GOE)\cite{Me91}:

\begin{equation}
\Sigma^2_{\rm GOE}(E) \approx \frac{2}{\pi^2}\left [ {\rm ln}(2\pi E)+
1.5772-\frac{\pi^2}{8} \right ] \; .
\label{GOE}
\end{equation}

\noindent
The logarithmic dependence on $E$ makes this value much smaller than the
one corresponding to uncorrelated spectra. The well-known spectral
rigidity of GOE follows: the number of levels in any energy interval is very
close to its mean value.

As said in the last part of the previous Section,
we routinely calculate the variance of the nearest-level spacing distribution
in order to check that the Wigner--Dyson value of 0.286 is recovered by our
numerical code. This is always the case for "reasonable" values of $W$ (see
Fig.\ref{fig.WD}) in spite of dealing with systems of finite size.
Notice that the calculation of $\Sigma^2(E)$ for very small values of $W$
is difficulted by finite size oscillations in the
density of states \cite{GL96}.

\begin{figure}
\begin{picture}(235,270) (-60,-60)
\epsfbox{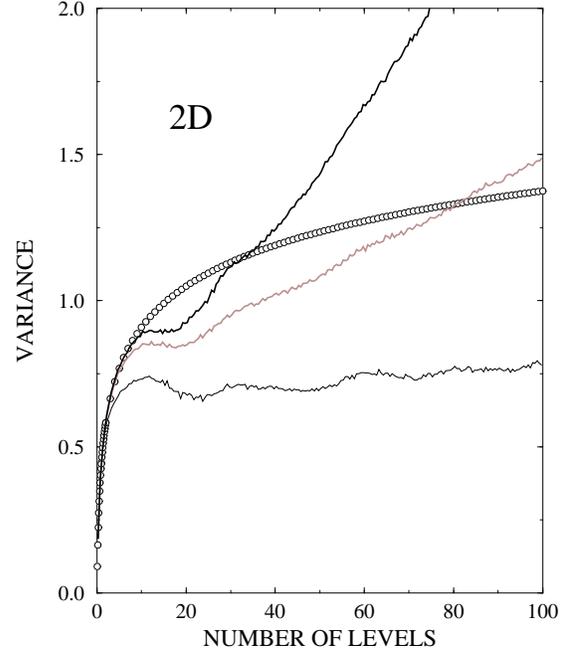}
\end{picture}
\caption{
$\Sigma^2(E)$ as a function of $E$ for the 2D billiard. The results
correspond to $L=32$ and disorder $W=2$ (black thin line),
$W=6$ (gray thick line), and $W=10$ (black thick line).
One thousand samples and all states in the region
$-3 \leq \epsilon \leq 3$ are included in the calculation.
Circles show the variance value corresponding to GOE fluctuations.
\label{fig.number_32x32}}
\end{figure}

Fig.\ref{fig.number_32x32} shows the typical behavior of the 
number variance $\Sigma^2(E)$ in different transport regimes.
Results correspond to the spectra of a large collection of
$32 \times 32$ clusters (one thousand samples). All states within the central
part of the band ($-3 \leq \epsilon \leq 3$) have been considered.
The black thin line corresponding to $W=2$ shows a typical
{\em ballistic} behavior:
fluctuations are really small in a large energy interval (the number of
levels fluctuates in less than one level for intervals containing 100 levels).
On the other hand, the black thick line shows a typical {\em diffusive} result
that corresponds to $W=10$. In this case, number fluctuation follows GOE
statistics for a small number of energy levels ($\approx 8$),
remains below GOE up to about 35 levels, and is well above
GOE values at larger energy scales.
In any case, fluctuations are still much smaller
than Poisson values ($\Sigma^2_{\rm Poisson}(100)=100$).
The gray thick line shows results corresponding to an intermediate
situation ($W=6$).
A natural physical explanation of such different behaviors
is provided by the ratio of the mean free path to the cluster side:
$l/L$ is about
0.7 for the diffusive system but 2.6 for the ballistic one. Although both
values are not too far from the accepted assumption $l \sim L$,
long--range energy correlations within the spectrum are quite different.
Our results for the intermediate value of $W$ prove that the transition
from one regime to the other is smooth.

Let us further comment our numerical result for bidimensional billiards
of side $L=32$ (Fig.\ref{fig.number_32x32}).
Firstly, a continuous increase of the number of levels following GOE
statistics is observed for increasing values of $W$ ($W=2, 6, 10$).
Secondly, it follows a region in which $\Sigma^2(E)$ deviates from GOE
showing a saturation effect. And finally, an almost linear increase of
the variance follows at larger energies.
Our qualitative interpretation of this behavior is the following.
Strict GOE statistics is limited by the maximum degeneracy of crystalline
bands. For a 2D cluster of the square lattice with open boundary conditions,
maximum degeneracy is 2 although splitting of the 8-fold levels corresponding
to periodic boundary conditions is really small. Surface roughness produces
a mixing of this degenerate states that is well described by the random
matrix elements of matrices of the GOE. This suffices to give Wigner-Dyson
statistics for the nearest-neighbor spacing distribution.
When larger energy scales are
considered, the fact that crystalline levels range from energy $-4$
to $4$ implies an incomplete mixing of levels (in the limit of very small
non--diagonal random coupling, $\Sigma^2(E)$ would be constant beyond the
number of levels corresponding to maximum degeneracy).
This explains saturation with variance values below GOE statistics.
The almost linear Poisson-like increase of the number variance at larger
energy is a subtle effect associated to the randomness of the energy levels at
the surface. Energy levels suffer uncorrelated shifts due to the formation
of surface resonances at uncorrelated energies. It is clear that for a large
enough value of $W$, about $4 L$ surface states appear outside the band.
For the same reason, about $4 L$ surface resonances exist within the band
for smaller $W$ values.
While saturation can be easily checked using matrices of random non--diagonal
elements but fixed diagonal elements, the Poisson-like linear increase
found at larger energies is a characteristic feature of our model.

\begin{figure}
\begin{picture}(235,270) (-60,-60)
\epsfbox{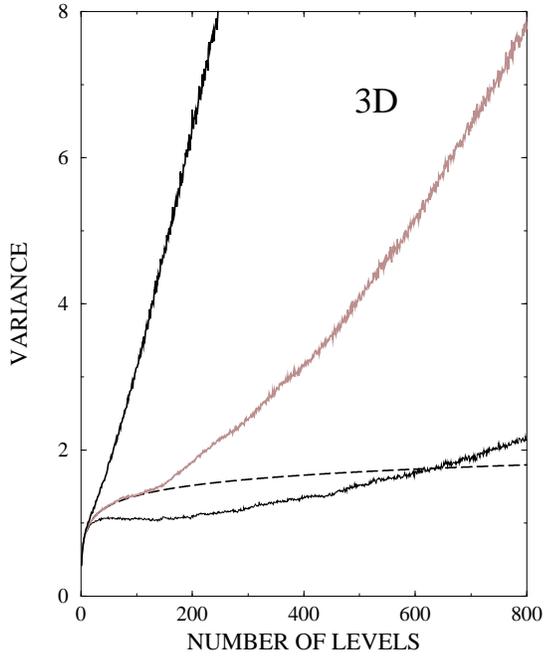}
\end{picture}
\caption{
$\Sigma^2(E)$ as a function of $E$ for the 3D billiard. The results
correspond to $L=20$ and disorder $W=2$ (black thin line),
$W=4$ (gray thick line), and $W=10$ (black thick line).
One hundred samples and all states in the region
$-4 \leq \epsilon \leq -2$ are included in the calculation.
The dashed line shows the variance value corresponding to GOE fluctuations.
\label{fig.number_20x20x20}}
\end{figure}

Similar results are obtained in 3D as shown by Fig.\ref{fig.number_20x20x20}.
A large collection (one hundred samples) of clusters with random surface
roughness has been fully diagonalized and the number variance has been 
calculated.
Fluctuations have been obtained for the central part of the band
excluding the band center: $-4 \leq \epsilon \leq -2$.
In this case, ballistic regime is clearly obtained for $W=2$
(black thin line) whereas
diffusive behavior is got for $W=10$ (black thick line).
An intermediate behavior is obtained for $W=4$ (gray thick line).
Results can be interpreted as in 2D. Strict GOE statistics is always
followed by at least six levels (typical 3D crsytalline degeneracy for
open boundary conditions). A region showing saturation of $\Sigma^2(E)$
follows (this is clearer for $W=2$ than in the other two cases).
At larger energy scales an uncorrelated behavior of the variance
can be seen. The larger slope of the linear region found for $W=10$ as
compared to $W=2$, reflects the fact that
surface resonances are better defined in the first case.
About $125$ energy levels closely follow GOE statistics for the
intermediate degree of surface roughness ($W=4$). We believe that this
property is related to a minimum value of the $l/L$ ratio. The smallest
value of the mean free path at fixed cluster size is obtained when
the mixing of crstalline states is maximum. Therefore, a larger number
of energy levels shows correlations that are well described by RMT.
Generally speaking, ballistic behavior ($l \geq L$) corresponds to a minimum
mixing of energy levels (only degenerate states are mixed by the surface
random perturbation) while diffusive behavior ($l \le L$) is seen when surface
resonances dominate the value of the number variance.

\begin{figure}
\begin{picture}(235,270) (-60,-60)
\epsfbox{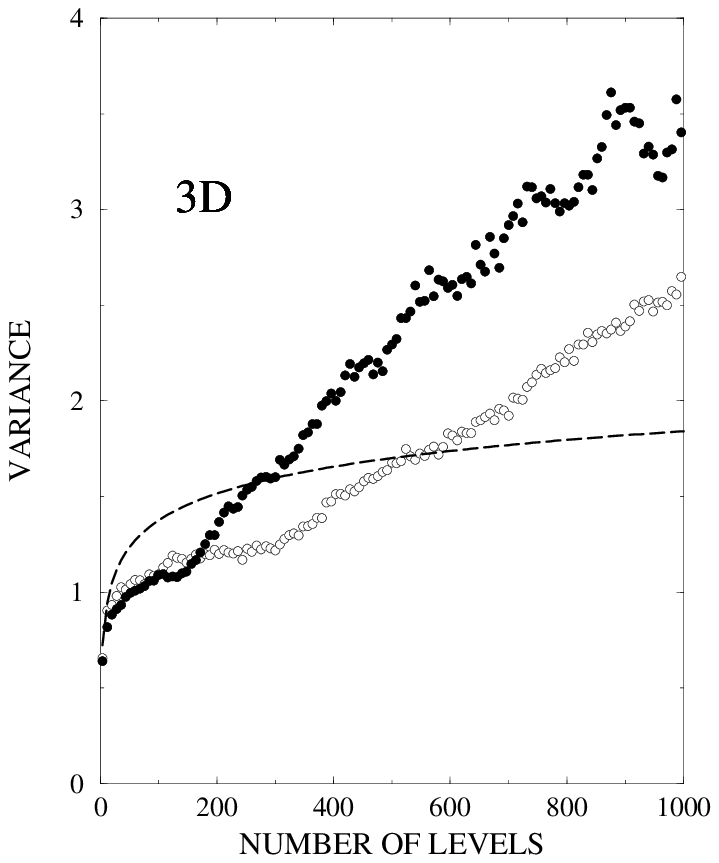}
\end{picture}
\caption{
Number variance as a function of the mean number of levels of a 3D
billiard of surface roughness given by $W=2$ and two different sizes.
Results for $L=14$ are given by filled circles whereas results for $L=20$
are given by empty circles. The corresponding GOE values are
shown by the dashed line.
\label{fig.number_W2}}
\end{figure}

Fig.\ref{fig.number_W2} shows the increase in the number of strongly
correlated energy levels for a fixed surface roughness ($W=2$) as the
size of the system increases. In this Figure, a clear discontinuous
change of slope can be seen for both curves. About 140 and 300 levels
are strongly correlated for $L=14$ and $20$, respectively, before the
effect of uncorrelated surface resonances dominates.
One could naively suggest an $L^2$ scaling for
the number of correlated levels. Nevertheless, a precise assesment about
the scaling of such critical energy would require further numerical
simulations well beyond our actual possibilities\cite{ec}.


\section{Concluding Remarks}

We have found that the mean free path in the model of
quantum chaotic billiards discussed in this work can be substantially smaller
than the system size. Actually, the ratio $l/L$ reaches a minimum value
of about 0.5 for the most relevant disorder values
($W$ of the order of the band width).
The transition from ballistic to diffusive behavior
is governed by the disorder parameter $W$ which simulates
surface roughness at a microscopic scale.
The behavior of the number variance statistics ($\Sigma^2(E)$ statistics)
clearly distinguishes between both regimes.
After closely following RMT for a small number of energy levels,
$\Sigma^2(E)$ shows a tendency to saturation before increasing
in an almost linear way at larger energies.
The region of saturation extends up to a really
large number of levels in the ballistic regime, 
while the linear increase quickly dominates in the diffusive regime.
Therefore, $\Sigma^2(E)$ is roughly below or above the values corresponding
to GOE statistics for ballistic or diffusive regimes, respectively.
In any case, both transport regimes
are still characterized by having a mean free path of the order of the
linear size of the system and $\Sigma^2(E)$ values much smaller than $E$, i.e.,
much smaller than values corresponding to completely uncorrelated energy
spectra. The behavior of the number variance has been related to the structure
of the particular type of random matrices produced by surface disorder.
Preliminary studies show that these results are not restricted to the
particular model analyzed in this work since similar results have been found
for a model with real surface roughness\cite{geometrico}.

\acknowledgments
This work was supported in part by the Spanish CICYT (grant MAT94-0058)
and DGICYT (grant PB93-1125). EC is grateful to  the ``Fundaci\'on Cultural
Privada Esteban Romero" for partial financial support.

\begin{references}
\bibitem{quasi}
Regions of small density of states favor the appearance of slightly
perturbed states because larger energy separations imply a diminution
of mixing among the states.
\end{references}
\end{document}